\documentstyle[preprint,aps,epsf,floats]{revtex}
\begin{document}
\tighten

\def\bfl{{\bbox \ell}}
\def\bull{\vrule height .9ex width .8ex depth -.1ex}
\def\Dslash{ {D\hskip-0.6em /} }
\def\MeV{{\rm MeV}}
\def\GeV{{\rm GeV}}
\def\Tr{{\rm Tr\,}}
\def\D{{\Delta}}
\def\Ds{{\Delta_6}}
\def\a{{\alpha}}
\def\b{{\beta}}
\def\c{{\gamma}}
\def\d{{\delta}}
\def\m{{\mu}}
\def\M{{\cal M}}
\def\slash{{\!\not\!}}
\def\nrcpt{NR\raise.4ex\hbox{$\chi$}PT\ }
\def\ket#1{\vert#1\rangle}
\def\bra#1{\langle#1\vert}
\def\ltap{\ \raise.3ex\hbox{$<$\kern-.75em\lower1ex\hbox{$\sim$}}\ }
\def\gtap{\ \raise.3ex\hbox{$>$\kern-.75em\lower1ex\hbox{$\sim$}}\ }
\newcommand{\gsim}{\raisebox{-0.7ex}{$\stackrel{\textstyle >}{\sim}$ }}
\newcommand{\lsim}{\raisebox{-0.7ex}{$\stackrel{\textstyle <}{\sim}$ }}
\def\grav{{NN\rightarrow NNh}}
\def\dila{{NN\rightarrow NN\phi}}
\def\NNT{{\cal A}}
\def\si{{}^1\kern-.14em S_0}
\def\siii{{}^3\kern-.14em S_1}

\def\Journal#1#2#3#4{{#1} {\bf #2}, #3 (#4)}

\def\NCA{\em Nuovo Cimento}
\def\NIM{\em Nucl. Instrum. Methods}
\def\NIMA{{\em Nucl. Instrum. Methods} A}
\def\NPB{Nucl. Phys. B}
\def\NPA{\em Nucl. Phys. A}
\def\PLB{Phys. Lett.  B}
\def\PRL{Phys. Rev. Lett.}
\def\PRD{Phys. Rev. D}
\def\PRC{Phys. Rev. C}
\def\PRA{{\em Phys. Rev.} A}
\def\PR{{\em Phys. Rev.} }
\def\ZPC{{\em Z. Phys.} C}
\def\PREP{{\em Phys. Rep.}  }
\def\ANN{{\em Ann. Phys.} }
\def\SCI{{\em Science} }
\def\CJP{{\em Can. J. Phys.}}

\preprint{\vbox{
\hbox{DOE/ER/40561-99-INT00}
\hbox{NT@UW-00-17}
}}
\bigskip
\bigskip

\title{Extra dimensions, SN1987a, and nucleon-nucleon scattering data}
\author{\bf Christoph Hanhart$^a$,  
Daniel R. Phillips$^a$~\footnote{Address after September 1: Department
of Physics, Ohio University, Athens, OH 45701},  
Sanjay Reddy$^b$, \\
 and
Martin J. Savage$^{a,c}$ }

\vspace{1cm}

\address{$^a$ Department of Physics 
\\ University of Washington, Seattle, WA 98195-1560}

\vspace{1cm}

\address{$^b$ Institute for Nuclear Theory
\\ University of Washington, Seattle, WA 98195-1550}

\vspace{1cm}

\address{$^c$ Jefferson Laboratory\\ 12000 Jefferson Avenue, Newport News, 
VA 23606}
\vspace{1cm}
\address{\tt hanhart, phillips, reddy and savage@phys.washington.edu}

\maketitle

\begin{abstract} 
One of the strongest constraints on the existence of large, compact,
``gravity-only" dimensions comes from SN1987a. If the rate of
energy loss into these putative extra dimensions is too high, then
the neutrino pulse from the supernova will differ from that actually
seen. The dominant mechanism for the production of
Kaluza-Klein gravitons and dilatons in 
the supernova is via gravistrahlung and dilastrahlung from the
nucleon-nucleon system. In this paper we compute the rates for these
processes in a model-independent way using low-energy theorems which
relate the emissivities to the measured nucleon-nucleon cross section.
This is possible because for soft gravitons and dilatons the leading
contribution to the energy-loss rate is from graphs in which the
gravitational radiation is produced from external nucleon
legs. Previous calculations neglected these mechanisms. We re-evaluate the
bounds on toroidally-compactified ``gravity-only" dimensions (GODs),
and find that consistency with the observed SN1987a neutrino signal
requires that if there are two such dimensions then
their radius must be less than 1 micron.  
\end{abstract}

\vfill\eject

\section{Introduction}

The conjecture that the four-dimensional universe we perceive is
merely a projection of a higher-dimensional space dates back at least
to the work of Kaluza and Klein in the 1930s. Such ideas imply that we
live in a world in which there are four infinite dimensions, and $n$
compactified ``extra" dimensions.  If this is the case there is the
possibility of observing Kaluza-Klein (KK) modes, which correspond to
excitations of ordinary standard-model particles in these extra
dimensions. The absence of any such effects in collider experiments
performed to date then places stringent limits on the size of such
extra dimensions, forcing them to have inverse radii larger than
$1~{\rm TeV}$~\cite{PDG}. Until ten years ago the conventional
wisdom was that the size of these extra dimensions would be set by
the Planck mass, and so would be much smaller than this.

In the last decade this conventional wisdom has been under attack from
two different, but connected, directions. Firstly, it has been
suggested that string theory could lead to compactified dimensions
which have radii much larger than $1/M_P$~\cite{An90,Wi96,Ly96}.
Secondly, an alternative picture of the extra dimensions has emerged,
in which the standard-model fields are confined to a four-dimensional
``brane", while gravity propagates in the $(n+4)$-dimensional
``bulk". This has generated considerable excitement because it seems
to provide a natural solution to the hierarchy problem. Effectively,
gravity is much weaker than the three other forces because it is
diluted via its propagation in the extra dimensions. One simple
scenario in which this occurs was suggested by Arkani-Hamed {\it et
al}.~\cite{ADD1,ADD2,AADD}. Suppose that there are $n$ compact
``gravity-only" dimensions (GODs) whose radius $R$ is large compared
to the scale of fundamental physics in the $(n+4)$-dimensional world,
$M_{(4+n)}$.  For distances $r \gg R$, the gravitational force will
behave like~\cite{ADD2}:

\begin{equation}
F=\frac{1}{M_{(4+n)}^{\, \, n+2}} \frac{1}{4 \pi R^n} \frac{m_1 m_2}{r^2},
\end{equation}
where $M_{(4+n)}$ is the fundamental scale of gravity in the 
$(n+4)$-dimensional
world. Matching to the Newtonian law of gravity, which is the effective 
theory for long distances and weak fields, we see that
\begin{equation}
4 \pi G_N=\left(\frac{1}{R M_{(4+n)}}\right)^n \frac{1}{M_{(4+n)}^{\, \, 2}},
\label{eq:GNreln}
\end{equation}
and so if $R \gg 1/M_{(4+n)}$ then the effective four-dimensional
Planck scale $M_P \sim 1/\sqrt{G_N}$ will be much larger
than the underlying $(n+4)$-dimensional scale $M_{(4+n)}$.

Of course, at distances $r \sim R$ the inverse-square law begins to
break down. Choosing $M_{(4+n)} \sim 1$ TeV implies, for a given $n$, 
a definite value of the scale $R$ at which this
breakdown occurs. In the case $n=1$ gravity
would be modified on solar-system distance scales,
but for $n=2$ the distance $R$ is already smaller than 1 mm. Thus, 
while the case of one GOD is ruled out experimentally, no
direct measurement of gravity has, as yet, been made that precludes
the existence of more than one GOD~\cite{ADD2}\footnote{We do not
consider Randall and Sundrum GODs~\cite{RS}.}.

An alternative to measuring the gravitational force
at distances of order one millimetre is to perform
collider physics searches at the scale $M_{(4+n)}$ and look for
``physics beyond the standard model"~\cite{HLZ,GRW,He}.  It is not
immediately apparent that such searches will reveal any
phenomenologically-significant effects. The only modification to
physics at the scale $M_{(4+n)}$ is in the physics of gravity, and so
the violations of the standard model appearing at that scale are,
naively, only of gravitational strength.  However, at such energies a
large number of graviton KK modes are excited, even though all
standard-model fields continue to behave as four-dimensional
objects. This compensates for the weak coupling between these KK modes
and matter fields.  Production of KK-gravitons, $h$, and KK-dilatons,
$\phi$, as well as their virtual contributions lead to deviations from
the predictions of the standard model, as demonstrated in
Refs.~\cite{HLZ,GRW,He}.  It is possible that such deviations could be
detected at future colliders~\cite{fc,colliderKK}.

Significant constraints have already been placed on the size of
compact GODs from precision experiments at existing
colliders~\cite{Lthree}. However, these collider bounds are not the
most stringent ones on the size of these dimensions.  Arkani-Hamed
{\it et al.}  pointed out that one of the strongest bounds on this
physics comes from SN1987a~\cite{ADD2}. The argument is very similar
to that used to bound the axion-nucleon coupling
strength~\cite{axionpapers,BT,BBT,Raffelt}. The ``standard model" of
supernovae does an exceptionally good job of predicting the duration
and shape of the neutrino pulse from SN1987a. Any mechanism which
leads to significant energy-loss from the core of the supernova
immediately after bounce will produce a very different neutrino-pulse
shape, and so will destroy this agreement---as demonstrated explicitly
in the axion case by Burrows, Brinkmann, and Turner~\cite{BBT}.
Raffelt has proposed a simple analytic criterion based on detailed
supernova simulations~\cite{Raffelt}: if any energy-loss mechanism has
an emissivity greater than $10^{19}$ ergs/g/s then it will remove
sufficient energy from the explosion to invalidate the current
understanding of Type-II supernovae's neutrino signal.  In the case of
interest here the bremsstrahlung processes $\grav$ (KK-gravistrahlung)
and $\dila$ (KK-dilastrahlung), provide an energy-loss mechanism for
the supernova.  Energy generated during the collapse is radiated away
into the extra-dimensional degrees of freedom, never to
return. Demanding that such reactions do not result in the loss of too
much energy from the supernova allows bounds to be placed on the size
of the toroidally-compactified GODs.

Two attempts to place such a constraint have already appeared in the
literature~\cite{CP,BHKZ}. Both assumed that the $NN$ scattering
amplitude was dominated by a single virtual pion exchange---an
uncontrolled approximation which we refer to hereafter as ``the
one-pion-exchange approximation". Although this approximation appears
reasonable, since one-pion exchange is the long-distance component of
the nuclear force, in fact the nuclear physics of this problem is much
more complicated. Most obviously, the very existence of nuclei tells
us that the strong nuclear force is non-perturbative. Yet, the
calculations of Refs.~\cite{CP,BHKZ} are only valid if one-pion
exchange can be treated in Born approximation.  Hence, the $NN$
interactions which occur during these radiative processes should,
instead, be described by the full $NN$ t-matrix.

One might think that such a calculation is model-dependent and
complicated. However, in the limit that the emitted radiation is soft
compared to the incoming $NN$ energy, the calculation simplifies
dramatically. In this limit there are low-energy theorems for the
emission of radiation which are analogous to those for photon
bremsstrahlung~\cite{Low,AD}. These theorems relate the amplitude for the
bremsstrahlung process to the on-shell $NN$ amplitude.  It thus
becomes possible to use $NN$ scattering data to calculate low-energy
emission processes directly. Recently, three of the present authors
used theorems of this sort for axial bremsstrahlung to perform a more
reliable calculation of axion and neutrino bremsstrahlung in a nucleon
gas~\cite{HPR}. The resultant rates are model-independent, and rely
only on two assumptions: that the radiation is soft, and that the
emission rate from the nuclear gas is dominated by two-body
collisions. The use of such an approach therefore essentially
eliminates the two-body dynamics as a source of uncertainty in the
problem.

In this paper we will derive model-independent low-energy theorems
which relate the amplitudes for $\grav$ and $\dila$ to the on-shell
$NN$ scattering amplitude. It turns out that at both leading and
next-to-leading order (LO and NLO) in the
soft expansion the cross sections for these production processes can
be expressed directly in terms of the $NN$ differential
cross section. We use the NLO production cross sections to
compute the energy loss from a supernova due to KK-graviton and KK-dilaton
radiation from the two-nucleon system. The ``Raffelt criterion"
explained above then leads directly to a constraint on the size of the
extra dimensions.

The paper proceeds as follows. In Section~\ref{sec-gravdila} we
derive the amplitude for the emission of soft gravitational radiation
from the $NN$ system. We include all terms which contribute up to
the first two orders in the soft expansion. We then take the
amplitude at NLO in the soft expansion and use it to
compute the energy loss due to the three processes $\grav$, $\dila$,
and $NN \rightarrow NNg$, where $g$ is the four-dimensional
graviton. In Section~\ref{sec-emiss} we derive formulae, meant for
numerical evaluation, which yield 
the emissivity of a gas of nucleons
due to these production mechanisms, give our ``best" result
for the bound on the size of the extra 
dimensions\footnote{In general what is actually constrained is the
$n$th root of the product $R_1 R_2 \ldots R_n$, where
$R_i$ is the radius of the $i$th extra dimension.}, and discuss
the validity of the soft approximation. Section~\ref{sec-analytic}
then examines the limiting cases of non-degenerate and degenerate
neutron matter.  In these limits analytic formulae for the emissivity
in terms of a single $NN$ cross section may be obtained. We also
exhibit the analytic formulae found if one assumes that
effective-range theory describes the $NN$ cross
section, and present the radius bounds given by these analytic
calculations. Finally, Section~\ref{sec-discuss} presents the results for the
supernova emissivity obtained with all of these different approaches,
and compares and contrasts them with the work of Refs.~\cite{CP,BHKZ}. 

Many-body effects, although undoubtedly important, were not the focus
of Ref.~\cite{HPR}, and will not be part of our discussion here. While
the radiation from the two-body system can be computed in a
model-independent way, the evaluation of many-body effects is
necessarily model-dependent, since it relies on selective
re-summations of classes of diagrams in the many-body problem.  One
set of many-body effects which can modify the vacuum rates for
KK-graviton emission was considered in Ref.~\cite{Di00}. We look
forward to future work on how many-body physics affects the reactions
$\grav$ and $\dila$, but we have not focused on these issues
here, since our goal was to provide a firm, model-independent,
foundation on which future calculations of many-body corrections to
these processes can build.


\section{$\grav$ and $\dila$}

\label{sec-gravdila}

\subsection{Graviton and dilaton dynamics in $n+4$ dimensions}

For simplicity, we let the $n$ extra dimensions form an $n$-torus
of radius $R=L/(2\pi)$ in each direction.
The zero-modes of the $(n+4)$-dimensional graviton correspond to the 
massless graviton, 
$h_{\mu\nu}$,
the $(n-1)$ vector bosons $A_\mu$ (which decouple from the matter fields) and 
$n(n-1)/2$ scalar fields, $\phi_{\mu\nu}$.
The graviton and scalar field mode expansions have the form
\begin{eqnarray}
h^{\mu\nu}(x,\vec{y}^{\,}) & = & \sum_{\vec j} h^{\mu\nu,\vec j}(x) \ 
\exp\left( i { 2\pi {\vec j}\cdot {\vec y}\over L}\right), \qquad
\phi_{ab}(x,\vec{y}^{\,})  =  \sum_{\vec j} \phi^{\vec j}_{ab}(x)\ 
\exp\left( i { 2\pi {\vec j}\cdot {\vec y}\over L}\right),
\label{eq:modes}
\end{eqnarray}
where $x$ are the 4-dimensional coordinates, $\vec{y}$ are the 
$n$-dimensional coordinates, and $\vec{j}$ 
is an $n$-dimensional vector representing
the momentum of the mode in the extra dimensions.
Although these modes are massless in the $(n+4)$-dimensional
world this extra-dimensional momentum results in an effective mass
on the four-dimensional brane, $m_{\vec j}$, with\footnote{Note
that this predicts the existence of a massless dilaton. This is unphysical,
since any mechanism that stabilizes the GODs at size $R$ will
provide a mass to the field $\phi_{ab}^{0}$. However, as long
as this mass is much smaller than the temperature of the nucleon 
gas in the supernova its precise value will not affect our calculation.}

\begin{equation}
m_{\vec j}^2  =  {4\pi^2|{\vec j}|^2 \over L^2}.
\label{eq:KKmass}
\end{equation}

We use KK-graviton fields with polarization
tensors $e^{\vec{j}}_{\mu \nu}(\vec{k},\lambda)$, where the
extra-dimensional momentum is $\vec{j}$, while the graviton is
propagating in the direction $\hat{k}$ in our world, and has
polarization $\lambda$. Then
\begin{eqnarray}
\sum_{\lambda=1}^5\ e^{\vec{j}}_{\mu \nu}(\vec{k},\lambda)
\left[e^{\vec{j}}_{\rho \sigma}(\vec{k},\lambda)\right]^*
\ =\ 
{1\over 2} B^{\vec{j}}_{\mu\nu ,\rho\sigma} (\vec{k})
,
\end{eqnarray}
with~\cite{HLZ,GRW}
\begin{eqnarray}
B^{\vec{j}}_{\alpha \beta \mu \nu}=\left(\eta_{\mu \alpha} \eta_{\nu \beta}
+ \eta_{\mu \beta} \eta_{\nu \alpha} - \eta_{\nu \mu} \eta_{\alpha \beta}
\right) &-&  \frac{1}{m_{\vec{j}}^2}\left(\eta_{\mu \alpha} k_{\nu} k_{\beta}
+ \eta_{\nu \beta} k_\mu k_\alpha + \eta_{\mu \beta} k_\nu k_\alpha
+ \eta_{\nu \alpha} k_\mu k_\beta\right)\nonumber\\
&+& \frac{1}{3}\left( \eta_{\mu \nu} + \frac{2}{m_{\vec{j}}^2}
k_\mu k_\nu \right)
\left( \eta_{\alpha \beta} + \frac{2}{m_{\vec{j}}^2} k_\alpha k_\beta \right).
\label{eq:Beqn}
\end{eqnarray}

The Lagrange density that describes the four-dimensional interactions between
the KK-graviton and KK-dilaton, and matter fields is then~\cite{HLZ,CP}
\begin{eqnarray}
{\cal L} & = & 
-{\kappa\over 2} \sum_{\vec j}
\left[\ 
h^{\mu\nu,\vec j} T_{\mu\nu}
 \ +\ \sqrt{2 \over 3(n+2)} \phi^{\vec j} T^\mu_\mu
\ \right],
\label{eq:lag}
\end{eqnarray}
where $T_{\mu\nu}$ is the conserved energy-momentum tensor
of the matter on the 4-dimensional brane, and 
the dilaton field is  $\phi=\phi_{aa}$. We have also 
defined $\kappa=\sqrt{32\pi G_N}$,
with $G_N$ being Newton's gravitational 
constant\footnote{Our $\kappa$ is a factor $\sqrt{2}$ larger
than that employed in Ref.~\cite{HLZ}, 
and our graviton polarization tensor is a factor of 2 smaller. This
change does not affect the answer computed for any observable.}.

\subsection{Radiation from the $NN$ system in the soft limit}

To compute gravitational radiation from the $NN$ system we
must calculate the energy-momentum tensor of free nucleons.
Neglecting isospin violation in the strong interaction, 
and electroweak effects
\begin{eqnarray}
T_{\mu\nu}^N & = & {i\over 2} \overline{\psi} 
\left( \overrightarrow\partial_\mu - \overleftarrow\partial_\mu\right)
\gamma_\nu\psi - g_{\mu\nu} {\cal L}
\ \rightarrow\ 
-{1\over 4M }  \overline{\psi} \left(\overrightarrow\partial_\mu 
- \overleftarrow\partial_\mu \right)\left(
 \overrightarrow\partial_\nu - \overleftarrow\partial_\nu
\right) \psi \ +\ ...
\ \ \ ,
\label{eq:EP}
\end{eqnarray}
where the ellipses denote spin-dependent interactions that do not
contribute if the momentum of the emitted radiation is small,
and terms that can
be made to vanish by the nucleon equations of motion.
$M$ is the isospin-averaged nucleon mass.
This, together with Eq.~(\ref{eq:lag}), defines 
the leading interaction between 
a single KK-graviton or KK-dilaton and a nucleon. 

%
\begin{figure}[t]
\centerline{{\epsfxsize=5.0in \epsfbox{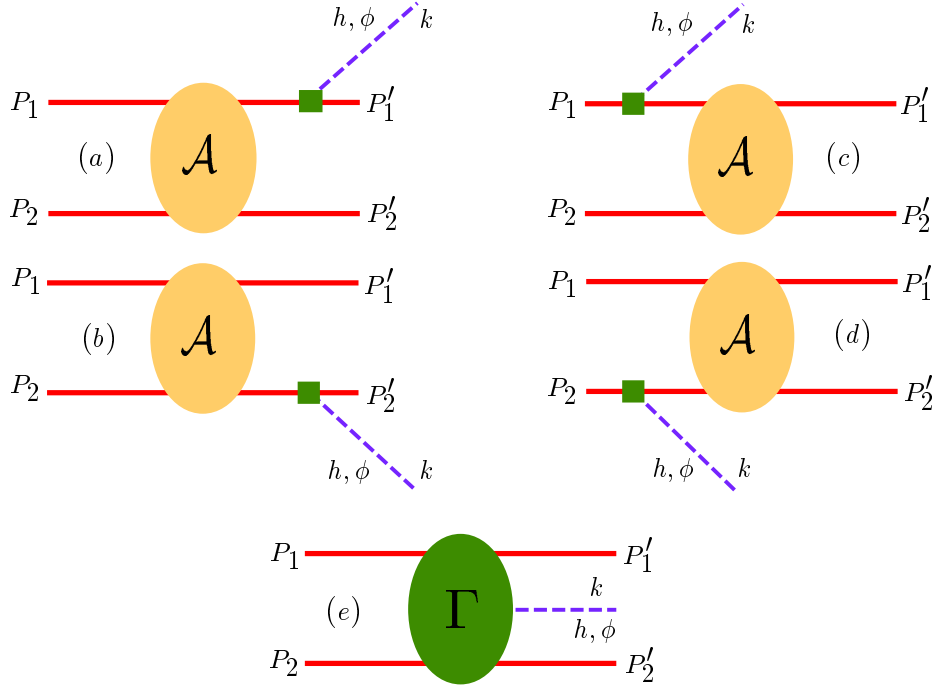}} }
\noindent
\caption{\it
The leading diagrams contributing to processes 
$\grav$ and $\dila$.
Nucleons are denoted by solid lines and the  KK-modes $h$ or $\phi$
are denoted by dashed lines.
Solid squares denote an insertion of the single-nucleon
energy-momentum tensor, while 
solid ovals containing ${\cal A}$ denote an insertion of the 
full $NN$ scattering amplitude.
The solid oval containing $\Gamma$ denotes the non-pole vertex 
required for the sum of diagrams to satisfy 
$\partial_\mu M^{\mu\nu}=0$.
}
\label{fig-brem}
\vskip .2in
\end{figure}

Now we fix our attention on the case of KK-graviton emission, and
examine the limit when the emitted graviton is
soft---which is to say its energy,
$\omega$, is much less than all other scales in the 
problem. This means that we certainly want $\omega \ll p^2/M$, 
the incoming energy of the $NN$ system, although other
restrictions will also emerge, as we shall see below. In what
follows we shall treat $\omega M/p^2$ as a small, dimensionless,
expansion parameter, and compute the contribution of the first two
orders in this expansion to the amplitude for $\grav$.

In the limit $\omega M/p^2 \ll 1$ the amplitude for gravistrahlung is
dominated by graphs (a)--(d) in Fig.~\ref{fig-brem}. In Fig.~\ref{fig-brem} 
the incoming nucleon momenta are $p_1$, $p_2$, the outgoing nucleon
momenta are $p_1'$, $p_2'$, and the KK-graviton four-momentum is denoted
by $k$. We must also define Mandelstam variables $s_i$ and $s_f$ for
the $NN$ interaction, depending on whether the nucleons scatter before
or after the graviton emission.  Likewise, two different Mandelstam
$t$'s enter, depending on whether the KK-graviton is emitted from nucleon
1 or nucleon 2. Denoting \begin{eqnarray} s_i=(p_1 + p_2)^2; \qquad
s_f=(p_1' + p_2')^2 \\ t_1=(p_1-p_1')^2; \qquad t_2=(p_2' - p_2)^2.
\end{eqnarray} 
The amplitudes for KK-graviton emission from the four
graphs (a)--(d) may then be written as:
\begin{eqnarray}
i M_{\mu \nu}^{(a)}&=&
\left(\frac{-i \kappa}{8 M}\right)
(2 p_1' + k)_\mu (2 p_1' + k)_\nu 
\bar{u}(\vec{p_1}') \bar{u}(\vec{p_2}')
\frac{i}{\not \! p_1'  + \not \! k - M}
(- i) {\cal A}(s_i,t_2) u(\vec{p_1}) u(\vec{p_2}),
\label{eq:Ma}\\
i M_{\mu \nu}^{(b)}&=&
\left(\frac{-i \kappa}{8 M}\right)
(2 p_2' + k)_\mu (2 p_2' + k)_\nu 
\bar{u}(\vec{p_1}') \bar{u}(\vec{p_2}')
\frac{i}{\not \! p_2'  + \not \! k - M}
(- i) {\cal A}(s_i,t_1) u(\vec{p_1}) u(\vec{p_2}),\\
i M_{\mu \nu}^{(c)}&=&
\left(\frac{-i \kappa}{8 M}\right)
\bar{u}(\vec{p_1}') \bar{u}(\vec{p_2}')
(- i) {\cal A}(s_f,t_2) 
\frac{i}{\not \! p_1 - \not \! k - M}
(2 p_1 - k)_\mu (2 p_1 - k)_\nu 
u(\vec{p_1}) u(\vec{p_2}),\\
i M_{\mu \nu}^{(d)}&=&
\left(\frac{-i \kappa}{8 M}\right)
\bar{u}(\vec{p_1}') \bar{u}(\vec{p_2}')
(- i) {\cal A}(s_f,t_1) 
\frac{i}{\not \! p_2 - \not \! k - M}
(2 p_2 - k)_\mu (2 p_2 - k)_\nu 
u(\vec{p_1}) u(\vec{p_2}).
\label{eq:Md}
\end{eqnarray}
This amplitude begins at $O(\chi^{-1})$, with the small parameter 
$\chi$ is meant as a mnemonic for any of the quantities 
$|\vec{k}|$, $\omega$, and $m_{\vec j}$. The $u$'s are 
Dirac spinors, normalized so that $\bar{u} u=1$.

Note that here the amplitude ${\cal A}$ is the on-shell $NN$ amplitude
projected onto positive-energy $NN$ states.  One might think that is
necessary to include in the expressions (\ref{eq:Ma})--(\ref{eq:Md})
the fact that one leg in the $NN$ interaction is off its mass shell.
However, it is clear that the difference between using the on-shell
$NN$ amplitude and the one that is ``correct" for the kinematics of
Fig.~\ref{fig-brem} is an $O(\chi^0)$ effect. The low-energy theorem
of Low~\cite{Low} for $NN \rightarrow NN \gamma$, and that of Adler
and Dothan~\cite{AD} for axial bremsstrahlung, are both based upon the
idea that we can construct the leading-order bremsstrahlung amplitude
using the on-shell $NN$ amplitude, and then determine the terms of
$O(\chi^0)$ using current conservation.  As we shall see below, a
similar low-energy theorem exists in this case, and so we need not
bother to include the off-mass-shell dependence of the $NN$ amplitude
in Eqs.~(\ref{eq:Ma})--(\ref{eq:Md}).  In fact, in general the off-shell
dependence of the $NN$ amplitude is the result of a particular choice
for the nucleon interpolating field, and so it can always be changed
by field redefinitions.  S-matrix elements are unaffected by such
field transformations~\cite{Haag}.

We now simplify the above expression, by using the fact that 
${\cal A}$ is projected onto positive-energy states, and
retaining  only the spin-independent terms up to
$O(\chi^0)$. Although it is possible to keep track of the
spin-dependent terms, we will drop them here, since they
always appear at $O(\chi^0)$, and so, after summing over nucleon spins,
do not make a contribution to the production cross section
until $O(\chi^0)$. This is two powers of $\chi$ beyond the leading term
in the cross section. Thus, excluding these terms, the result is:
\begin{eqnarray}
M_{\mu \nu}^{(a)}&=& -\frac{\kappa}{4}
(2 p_1' + k)_\mu (2 p_1' + k)_\nu 
\frac{1}{2 p_1' \cdot k + k^2} \left(1 + \frac{1}{2M^2} 
k \cdot p_1'\right) {\cal A}(s_i,t_2);\\
M_{\mu \nu}^{(b)}&=& - \frac{\kappa}{4}
(2 p_2' + k)_\mu (2 p_2' + k)_\nu 
\frac{1}{2 p_2' \cdot k + k^2} \left(1 + \frac{1}{2M^2} 
k \cdot p_2'\right) {\cal A}(s_i,t_1); \\
M_{\mu \nu}^{(c)}&=& \frac{\kappa}{4} 
{\cal A}(s_f,t_2) \left(1 - 
\frac{1}{2 M^2} k \cdot p_1\right) 
\frac{1}{2 p_1 \cdot k - k^2}
(2 p_1 - k)_\mu (2 p_1 - k)_\nu;\\
M_{\mu \nu}^{(d)}&=& \frac{\kappa}{4}
{\cal A}(s_f,t_1) \left(1 - \frac{1}{2 M^2} k \cdot p_2 \right)
\frac{1}{2 p_2 \cdot k - k^2} (2 p_2 - k)_\mu (2 p_2 - k)_\nu,
\end{eqnarray}
where we have also used the fact that the external nucleon legs are on-shell.

\subsection{Conserved $M_{\mu \nu}$}

Observe that if we sum graphs (a)--(d) to get $M_{\mu \nu}$ then 
the result is not a conserved quantity. 
To remedy this, we generalize the prescription
used in photon bremsstrahlung~\cite{Low,AD}, writing:

\begin{equation}
k^\mu M_{\mu \nu}=-k^\mu \Gamma_{\mu \nu} + O(\chi^2),
\end{equation}
and then construct

\begin{equation}
M^{(0)}_{\mu \nu}=M_{\mu \nu} + \Gamma_{\mu \nu}.
\end{equation}
By the same arguments that apply in the case of the Low~\cite{Low} and
Adler-Dothan~\cite{AD} theorems, $M_{\mu \nu}^{(0)}$
will be the correct amplitude up to
spin-dependent pieces of order $\chi^0$ and spin-independent pieces
of $O(\chi^1)$~\footnote{Structures involving the spin-matrix
$\sigma_{\mu \nu}$ which are not constrained by current conservation
can contribute to $M_{\mu \nu}$ at order $\chi^0$. However, the effect
of such terms on the observables computed here is suppressed by two
powers of $\chi$ compared to the leading soft result.}.

A brief calculation then reveals that the spin-independent piece of
the amplitude for $NN \rightarrow NN h$ is, up to $O(\chi^0)$:
\begin{eqnarray}
M_{\mu \nu}^{(0)}&=&\frac{\kappa}{4} 
\left[{\cal A}(s_f,t_2) \frac{1}{2 p_1 \cdot k - 
k^2} \left(2 p_1 - k\right)_\mu \left(2p_1 - k\right)_\nu
+ {\cal A}(s_f,t_1) \frac{1}{2 p_2 \cdot k - k^2} 
\left(2 p_2 - k\right)_\mu 
\left(2 p_2 - k\right)_\nu \right. \nonumber\\
&-& \left(2 p_1' + k\right)_\mu \left(2 p_1' + 
k\right)_\nu 
\frac{1}{2 p_1' \cdot k + k^2} {\cal A}(s_i,t_2) 
- \left(2 p_2' + k\right)_\mu \left(2 p_2' + k\right)_\nu 
\frac{1}{2 p_2' \cdot k + k^2} {\cal A}(s_i,t_1) \nonumber\\
&+& 2 \eta_{\mu \nu} {\cal A}(\bar{s},\bar{t}) 
+ 4 (p_1 + p_2)_\mu (p_1 + p_2)_\nu
\left. \frac{\partial {\cal A}(s,\bar{t})}{\partial s} \right|_{s=\bar{s}}
+ \left. 4 (p_2 - p_2')_\mu (p_2 - p_2')_\nu 
\left. \frac{\partial {\cal A}(\bar{s},t)}{\partial t} \right|_{t=\bar{t}} 
\right].\nonumber\\
\label{eq:M0}
\end{eqnarray}

The classical amplitude for the emission of gravitational radiation from
a system of colliding particles was derived many years ago by 
Weinberg~\cite{Weinbergbook}. We may now recover this result by
retaining only the $O(\chi^{-1})$ parts of 
Eq.~(\ref{eq:M0}):

\begin{equation}
M_{\mu \nu}^{(-1)}
=\frac{\kappa}{2} \sum_j \eta_j {p_j}_\mu {p_j}_\nu
\frac{1}{p_j \cdot k} {\cal A}(\bar{s},\bar{t}),
\label{eq:Weinberg}
\end{equation}
where $j$, the particle index, runs from 1 to 4, 
with $j=1,2$ for incoming nucleons and $j=3,4$ 
for outgoing nucleons, and

\begin{eqnarray}
&& \quad \eta_j=\left\{ \begin{array}{cl}
               1  & \hbox{if $j=1,2$}\\
               -1 & \hbox{if $j=3,4$};
               \end{array}
       \right.\\
&& \bar{s}=\frac{s_i + s_f}{2}; \quad \bar{t}=\frac{t_1 + t_2}{2}.
\end{eqnarray}
Weinberg's derivation of Eq.~(\ref{eq:Weinberg})
is entirely
classical, which is not surprising, since the graphs that contribute
to the amplitude only involve radiation
from external legs, and so they correspond to classical
bremsstrahlung from the gravitational charges accelerated during
the $NN$ collision. It is also worth pointing out that the Weinberg
result is only valid if the $NN$ t-matrix is slowly varying 
over the energy region of interest\footnote{Weinberg's basic result
actually omits the amplitude ${\cal A}$. He assumes the
scattering that leads to the emission of the radiation is isotropic, and
then includes anisotropy in the two-body scattering by multiplying by the
two-particle differential cross section.}.

Next observe that the conservation of $M$---$k^\mu M_{\mu \nu}=0$---implies
the following relations:

\begin{eqnarray}
M_{0 i}&=&-\frac{k^j}{\omega} M_{ji},\\
M_{0 0}&=&\frac{k^i k^j}{\omega^2} M_{ji}.
\end{eqnarray}
This obviates the necessity to actually calculate these ``0" 
components. These relations may be formalized by defining the symbol 
$W_{\mu \nu i j}$ as follows:

\begin{equation}
W_{\mu \nu i j}=\eta_{i \mu}\eta_{j \nu} - \eta_{i \mu}\eta_{0 \nu} 
\frac{k_j}{\omega} - \eta_{0 \mu} \eta_{j \nu} \frac{k_i}{\omega}
+ \eta_{0 \mu} \eta_{0 \nu} \frac{k_i k_j}{\omega^2}.
\label{eq:Weinsymbol}
\end{equation}
If $M_{\mu \nu}$ is conserved then we need only
calculate the space-space components of this tensor. Other portions can
be generated via the relation:

\begin{equation}
M_{\mu \nu}=W_{\mu \nu i j} M^{i j}.
\end{equation}
In particular, if we work in the centre-of-mass frame, 
then at leading order in the nucleon velocity
the space-space components of $M_{\mu \nu}$ have a very simple form:
\begin{equation}
M_{ij}^{(-1)}=
{\cal A} (\bar{s},\bar{t}) \frac{\kappa}{M \omega}
\left[p_i p_j - p_i' p_j'\right].
\label{eq:LOM}
\end{equation}
This result differs from Eq.~(\ref{eq:M0}) by relativistic 
corrections, which are suppressed by two powers
of $p/M$. In addition, the next-to-leading order (NLO) 
amplitude (\ref{eq:M0}) contains
corrections to Eq.~(\ref{eq:LOM}) which are  
suppressed by $\omega/M$. Since $\omega \ll p^2/M$, by
assumption, the expansion in $p/M$ is less convergent than the one
in $\omega/M$. 

If we continue to neglect these effects of order $p^2/M^2$ and
$\omega/M$, then we can also write down a compact formula for the
next-order correction to $M_{ij}$:
\begin{equation}
M_{ij}^{(0)}=\kappa \left[-\frac{1}{2} \delta_{ij}{\cal A} (\bar{s},\bar{t}) 
+ (p_i - p_i') (p_j - p_j')
\left. \frac{\partial {\cal A}}{\partial t}(\bar{s},t)\right|_{t=\bar{t}}
\right],
\label{eq:NLOM}
\end{equation}
where, once again, we have chosen to work in the centre-of-mass frame.

Note that working only at leading order involves neglecting the pieces
of the amplitude $M_{\mu \nu}$ corresponding to variations of the $NN$
t-matrix in the Mandelstam variables $s$ and $t$. This is clearly not
a good approximation at low energies, where the $NN$ t-matrix varies
rapidly.  At higher energies the scale of variation of ${\cal A}$ can
be thought of as $m_\pi$, the pion mass, and so we would also like to
have $\omega \ll m_\pi$. In fact, this is probably a rather
conservative estimate, since empirically the energy and
angle-dependence of the $NN$ cross section is quite weak at these
higher energies. However, a proper test of the convergence of the
soft expansion is really required.

One obvious way to perform such a test is to 
compute the LO result (\ref{eq:LOM}) and then compare it to the result
at NLO. In Section~\ref{sec-NLO} we will show that in fact
the NLO correction vanishes, and so the LO amplitude
(\ref{eq:LOM}) is more reliable than one might naively expect.

\subsection{Result for energy loss from the $NN$ system to leading order}

\label{sec-LO}

In this Section we will compute the energy-loss for a given graviton
KK-mode $\vec{j}$.  In Section~\ref{sec-emiss} we sum over all modes
in order to compute the total rate of energy emission from a gas of
nucleons.

The energy loss due to graviton emission into a given KK mode $\vec{j}$ 
is given by:

\begin{equation}
d \varepsilon_h^{\vec{j}}=\omega \frac{d^3 k}{(2 \pi)^3 2 \omega} 
\sum_\lambda \left[{M}^{\alpha \beta} 
e^{\vec{j}}_{\alpha \beta}(\vec{k},\lambda)\right]^* 
e^{\vec{j}}_{\mu \nu}(\vec{k},\lambda) {M}^{\mu \nu}.
\label{eq:desire}
\end{equation}
(Strictly speaking this is not an energy loss, since we need
to multiply by the density of initial and final $NN$ states, however, 
throughout this section we employ this somewhat lax terminology.)

Using the symbol $W$, defined in Eq.~(\ref{eq:Weinsymbol}), we
may rewrite this as
\begin{equation}
d\varepsilon_h^{\vec{j}}=
\frac{dk \, k^2}{4 \pi^2} \frac{d \hat{k}}{4 \pi} \frac{1}{2}
M_{ij} W^{\mu \nu i j} B^{\vec j}_{\mu \nu \alpha \beta} 
W^{\alpha \beta k l} M_{kl},
\end{equation}
where $B$ is defined by Eq.~(\ref{eq:Beqn}).
Since $k^\mu W_{\mu \nu k l}=0$ by construction, only the pieces of
$B$ proportional to $\eta$ symbols survive. After some calculation we see that:
\begin{eqnarray}
&& \frac{1}{2} W_{\mu \nu i j} B^{\mu \nu \alpha \beta} W_{\alpha \beta k l}
=\frac{1}{2} \left(\eta_{ik} \eta_{jl} + \eta_{il} \eta_{jk}
- \frac{2}{3} \eta_{ij} \eta_{kl}\right)\nonumber\\
&& \qquad + \frac{1}{2 \omega^2} \left(\eta_{ik} k_j k_l 
+ \eta_{jk} k_i k_l + \eta_{il} k_j k_k + \eta_{jl} k_i k_k
- \frac{2}{3} (k_i k_j \eta_{kl} + k_k k_l \eta_{ij})\right)
+ \frac{2}{3 \omega^4} k_i k_j k_k k_l.\nonumber\\
\end{eqnarray}

At this point we may integrate over the final-state graviton direction
$\hat{k}$, since at LO $M_{ij}$ does not depend on $\hat{k}$. 
This gives:
\begin{eqnarray}
&& \frac{1}{2} \int \frac{d \hat{k}}{4 \pi} \, W_{\mu \nu i j}
B^{\mu \nu \alpha \beta} W_{\alpha \beta k l}=
\frac{19}{90}(\delta_{ik} \delta_{jl} + \delta_{il} \delta_{jk})
- \frac{1}{15} \delta_{ij} \delta_{lk}\nonumber\\
&& \qquad + \frac{m_{\vec{j}}^2}{\omega^2}\left(\frac{11}{45}(\delta_{ik} 
\delta_{jl} + \delta_{il} \delta_{jk}) - \frac{14}{45}\delta_{ij} 
\delta_{kl}\right)
+ \frac{m_{\vec{j}}^4}{\omega^4} \frac{2}{45} (\delta_{ik} \delta_{jl}
+ \delta_{il} \delta_{jk} + \delta_{ij} \delta_{kl}).
\label{eq:angleavg}
\end{eqnarray}
Using the LO $M_{\mu \nu}$, Eq.~(\ref{eq:LOM}),
we finally obtain the following energy loss per unit frequency
interval, for radiation into a KK mode of given mass $m_{\vec{j}}$,
\begin{equation}
\frac{d\varepsilon_h}{d \omega}=\frac{k}{\omega} \frac{8 G_N}{5 \pi}
\frac{4 \overline{p^2}^2}{M^2} \sin^2 \theta_{\rm cm} |{\cal A}|^2
\left(\frac{19}{18} + \frac{11}{9} \frac{m_{\vec{j}}^2}{\omega^2}
+ \frac{2}{9} \frac{m_{\vec{j}}^4}{\omega^4}\right),
\label{eq:dEdomegaKK}
\end{equation}
where $\theta_{\rm cm}$ is the centre-of-mass scattering angle:
$\cos \theta_{\rm cm}=\hat{p} \cdot \hat{p'}$,
and $\overline{p^2}=(p^2 + {p'}^2)/2$.

This argument can also be pursued for 4-dimensional gravitons, 
provided one modifies the polarization tensor $B$ appropriately. The result
for energy-loss due to 4D-graviton emission is:
\begin{equation}
\frac{d\varepsilon_g}{d \omega}=\frac{8 G_N}{5 \pi} \frac{4 \overline{p^2}^2}
{M^2} \sin^2\theta_{\rm cm} |{\cal A}|^2,
\label{eq:dEdomegagravi}
\end{equation}
as derived by Weinberg~\cite{Weinbergbook}. Note that this is {\it not}
the $m_{\vec{j}} \rightarrow 0$
limit of Eq.~(\ref{eq:dEdomegaKK}), since the 4D-graviton has fewer degrees
of freedom, due to its 4D-masslessness. Note also that here, and 
in Eqs.~(\ref{eq:dEdomegaKK}) and (\ref{eq:dEdomegadila}), we have
made the replacement
\begin{equation}
p^4 + {p'}^4 - 2 p^2 {p'}^2 \cos^2 \theta_{\rm cm} \quad
\longrightarrow \quad 2 \overline{p^2}^2 \sin^2 \theta_{\rm cm},
\end{equation}
in order to make contact with Ref.~\cite{Weinbergbook} and simplify
the formulae. This corresponds to evaluating the energy of the $NN$
system at the point $\bar{s}$. The error introduced by this
replacement is of order $M^2 \omega^2$, and so is suppressed by two
powers of $\chi$.

Meanwhile, the result for energy loss due to KK-dilaton emission is also
obtained by an analogous argument. It is:
\begin{equation}
\frac{d\varepsilon_\phi}{d \omega}= \frac{k}{\omega} \frac{8 G_N}{5 \pi}
\frac{4 \overline{p^2}^2}{M^2} \sin^2 \theta_{\rm cm} |{\cal A}|^2
{2\over 9 (n+2)}
\left(1-\frac{m_{\vec{j}}^2}{\omega^2}
\right)^2.
\label{eq:dEdomegadila}
\end{equation}

\subsection{Next-to-leading order calculation}

\label{sec-NLO}

These energy-loss rates represent the leading-order result, which we
shall denote by $O(\chi^{-2})$.  To calculate the correction to this
energy-loss rate which arises at NLO we must contract the amplitude
(\ref{eq:LOM}) with the correction (\ref{eq:NLOM}). Since---at least
in the non-relativistic limit---the space-space components of the NLO
amplitude do not depend on the direction of the emitted radiation, we
may use the steps (\ref{eq:desire})--(\ref{eq:angleavg})
employed above to show that the piece of $\frac{d \varepsilon_h}{d
\omega}$ which arises at $O(\chi^{-1})$ vanishes.

To do this we first note that the trace of $M^{(-1)}$ over its
spatial components, $M^{(-1)}_{ii}$, is $O(\chi^0)$. Therefore, since
we are trying to compute pieces of $\frac{d \varepsilon_h}{d \omega}$
that are $O(\chi^{-1})$ we can neglect terms such as $M_{ii}
M_{jj}$. It also follows that the contraction of the $\delta_{ij}$
piece of $M^{(0)}_{ij}$ with $M^{(-1)}_{kl}$ only produces pieces of the
energy-loss rate that are of higher order than we are concerned with
here.

A brief calculation then shows that the contraction of
the piece of $M^{(0)}_{ij}$ proportional to 
$\frac{\partial {\cal A}}{\partial t}$ with $M^{(-1)}_{kl}$
also only yields pieces of $O(\chi^0)$. 
Thus, it follows that
the first corrections to the expressions (\ref{eq:dEdomegaKK}), 
(\ref{eq:dEdomegagravi}), and (\ref{eq:dEdomegadila}) are
suppressed by two powers of the small parameter $\chi$, which
we can now identify more precisely as
\begin{equation}
\chi=\frac{\omega M}{\overline{p^2}}.
\label{eq:chidefn}
\end{equation}

It is worthwhile stressing that the NLO correction to these
results would {\it not} be zero if we had used, say, the
initial-state energy $p^2/M$ in these formulae instead of the average
energy of the $NN$ state: $\overline{p^2}/M$.
It must also be remembered that at $O(\chi^0)$ there exist
contributions to the energy-loss rates which 
are not constrained by the dual requirements of the
on-shell $NN$ data and the conservation of $M_{\mu \nu}$. 

\section{Gravitational emissivity of a gas of nucleons}

\label{sec-emiss}

\subsection{Non-Relativistic Nucleons at Arbitrary Degeneracy}

The gravitational emissivity of a nucleon gas due to the two-body reaction
$a+b \rightarrow a+b+X$, where the labels $a$ and $b$ can 
represent either a neutron
or a proton, and $X$ is the type of gravitational radiation
(massless 4-D gravitons (g) or massive Kaluza-Klein gravitons 
(h) or dilatons ($\phi$)), is given by the general
formula
\begin{eqnarray}
{d{\cal E}_X \over dt} & = & \sum_{\vec j}
\int d \omega
\int \left[ \prod_{i=1..2} \frac{d^3p_i \, d^3p_i'}{(2\pi)^6} \right]
S~f_a(E_1)f_b(E_2)(1-f_a(E_1'))(1-f_b(E_2')) \nonumber \\
&&\qquad (2\pi)^4 \delta^4(p_1+p_2-p_1'-p_2'-k) \, 
\frac{d\varepsilon_X}{d \omega}
\label{eq:emissgen},
\end{eqnarray}
Here $\frac{d\varepsilon_X}{d\omega}$ is the energy-loss per unit
frequency due to $NN \rightarrow NN X$ computed above, summed over initial and
final spins, for the emission of KK-gravitons [see Eq.~(\ref{eq:dEdomegaKK})],
ordinary gravitons [see Eq.~(\ref{eq:dEdomegagravi})], or KK-dilatons
[see Eq.~(\ref{eq:dEdomegadila})]. Again, as noted above, 
Eqs.~(\ref{eq:dEdomegaKK})--(\ref{eq:dEdomegadila}) are not
really energy-loss rates, since they do not include the $NN$ density of
states. However, when integrated as in Eq.~(\ref{eq:emissgen}) they will
yield the emissivity. Meanwhile, $f_a$ and $f_b$ are the nucleon 
Fermi distribution functions for the species $a$ and $b$. 
The symmetry factor $S=[1/(1+\delta_{ab})]^2$ is different in the $np$ and
$nn$ cases. The sum over KK modes clearly does not apply to the case
when $X$ is the four-dimensional graviton.

The formula (\ref{eq:emissgen}) applies equally well
to the $np$ and $nn$ cases. Although we could proceed in complete
generality and consider a nucleon gas with different densities of neutrons
and protons, we will specialize our treatment at this point to the $nn$
case, since the formulae are more transparent there.
The proton fraction is generically small in supernovae, and so the inclusion
of $np$ scattering---which is easily done using steps analogous
to those discussed below---should make little difference to the radius bound
we obtain.

In the soft-radiation limit we may neglect the momentum $\vec{k}$
in the three-momentum conserving delta-function of Eq.~(\ref{eq:emissgen}).
Spherical symmetry and energy-momentum conservation can also be
exploited, thereby eliminating nine of the fifteen integrals appearing in
Eq.~(\ref{eq:emissgen}). Defining total, initial relative, and final relative 
momenta $P$, $p$, and $p'$ we move to dimensionless variables~\cite{BT}
\begin{equation}
u_P=\frac{P^2}{8 M T}; \qquad u_r=\frac{p^2}{2 M T}; \qquad 
u_r'=\frac{p'^2}{2MT},
\end{equation}
where $T$ is the temperature of the neutron gas.

Denoting $\cos(\theta)=\hat{p} \cdot \hat{P}$, $\cos(\theta')=\hat{p'} \cdot
 \hat{P}$
and $\phi$ to be the angle between $\vec{p}$ and $\vec{p'}$ projected onto
the $\hat{P}$ plane, we are left with a six-dimensional integral:
\begin{eqnarray}
&& {d{\cal E}_X \over dt} = \sum_{\vec j}
S \frac{2^{15/2} G_N M^{9/2} T^{13/2}}{5 \pi^6}
\int_{\delta_{\vec{j}}}^\infty \, du_r \int_{-1}^1 \, d(\cos \theta)
\int_0^\infty \, du_P \int_0^{u_r - \delta_{\vec{j}}} \, du_r' 
\int_{-1}^1 \, d(\cos \theta')  \nonumber\\
&& \qquad u_r^{1/2} u_P^{1/2} u_r'^{1/2} \bar{u}^2 \
\xi_X [\delta_{\vec{j}}/(u_r - u_r')] \
f_1 f_2 (1-f'_1) (1-f'_2) \int_0^{2 \pi} \frac{d \phi}{2 \pi} \,
\sin^2 \theta_{\rm cm} \,|{\cal A}(\theta_{\rm cm},2T\bar{u})|^2.
\nonumber\\
\label{eq:uemiss}
\end{eqnarray}
Here $f_i=[\exp(u_i-y)+1]^{-1}$, 
with $u_{1,2}=u_P + u_r \pm 2 \sqrt{u_P u_r} \cos \theta$,
and $y=\mu/T$. The $f'_i$s are defined similarly in the final state,
with $u'_{1,2}=u_r' + u_P \pm 2 \sqrt{u_r' u_P} \cos \theta'$.
We have also defined $\delta_{\vec{j}}=m_{\vec{j}}/(2T)$ and
$\bar{u}=(u_r + u_r')/2$, and written ${\cal A}$ as a function of the 
centre-of-mass
scattering angle and the centre-of-mass energy.
An implicit summation of $|{\cal A}|^2$ over both initial and final
nucleon spins is to be understood here, and throughout what follows.
The function $\xi$ is different depending on the species $X$ under 
consideration:
\begin{eqnarray}
\xi_g[x]&=&1,\\
\xi_h[x]&=&\sqrt{1 - x^2} \left(\frac{19}{18} + \frac{11}{9} x^2 + \frac{2}{9} 
x^4\right),\\
\xi_\phi[x]&=&\frac{2}{9(n+2)}\left(1 - x^2\right)^{5/2}.
\end{eqnarray}
The dependence of the integrand on the angle $\phi$ enters solely
through the centre-of-mass scattering angle $\theta_{\rm cm}$, which is
related to $\theta$, $\theta'$ and $\phi$ via
\begin{equation}
\cos{\theta_{cm}}=\cos \theta  \cos \theta' + \sin \theta \sin \theta'
\cos \phi\,.
\end{equation}

Now we wish to perform the sum over KK-graviton modes. If the modes are narrowly
spaced we may rewrite:
\begin{equation}
\sum_{\vec{j}} \quad \longrightarrow \quad
R^n \Omega_n \int dm \, m^{n-1},
\end{equation}
where $\Omega_n$ is the surface area of the $n$-dimensional sphere.
The order of integration may then be interchanged, and the integral
of the function $f$ over the graviton masses evaluated. We may also
change variables to $u_r'$ and $\alpha=\omega/T$, so that 
ultimately we obtain:
\begin{equation}
{d{\cal E}_X \over dt} = \frac{4 G_N}{3} (2 T)^{13/2} (T R)^n
\pi^{\frac{n+1}{2}} M^{1/2} h_n
\int_{0}^\infty \, d\alpha  \int_0^{\infty} du_r'
K^{(n)}(\alpha,u_r'),
\label{eq:full}
\end{equation}
where the dimensionless kernel $K^{(n)}$ is:
\begin{eqnarray}
&& K^{(n)}(\alpha, u_r')=\left(u_r' + \frac{\alpha}{2}\right)^{1/2} 
u_r'^{1/2} \alpha^n \left(u_r' + \frac{\alpha}{4}\right)^2
\frac{M^4}{(2 \pi)^6}
\int_{-1}^1 d(\cos \theta') \,
\int_{-1}^1 d(\cos \theta) \,
\nonumber\\
&& \qquad \int_0^{2 \pi} \frac{d \phi}{2 \pi}
\, \sin^2 \theta_{\rm cm} \, 
|{\cal A}(\theta_{\rm cm},T[2u_r' + \alpha/2])|^2
\int_0^\infty \, du_P \, \sqrt{u_P} f_1 f_2 (1-f'_1) (1-f'_2),
\label{eq:kernel}
\end{eqnarray}
and
\begin{equation}
h_n=\frac{3n^3 + 24n^2 + 55n + 42}{(n+2)(n+3)(n+5)}
\frac{1}{\Gamma\left(\frac{n+3}{2}\right)},
\end{equation}
if we consider the radiation of KK-gravitons and KK-dilatons.
Note that the gravitons alone give 
\begin{equation}
h_n=\frac{3n^2 + 18n + 19}{(n+5)(n+3)}
\frac{1}{\Gamma\left(\frac{n+3}{2}\right)},
\end{equation}
which, for $n=2$, differs from the full result by about 1.5\%. 

Note that the number of GODs, $n$, enters here through the power of
the radiated energy which appears in the kernel, and through the
pre-factor. Different models of the extra-dimensional physics would
change the density of states in the integral over the KK-graviton
modes, and so would give a different overall pre-factor, but the
structure of the answer for the emissivity would be essentially the
same. The crucial fact about the extra dimensions
which makes the emissivity (\ref{eq:full}) sizable during SN1987a
is that for $R \sim 1~{\rm mm}$ the dimensionless quantity $TR$ is of
order $10^{11}$, and so many KK modes are excited during the $NN$ collisions
that take place in the violent supernova environment.

\subsection{Numerical results and bounds on the size of extra dimensions}

During the supernova, the newly-born neutron star (proto-neutron star) 
has a typical temperature of 40--60 MeV and a
high baryon density corresponding to a neutron chemical potential
$\mu_n \simeq 50-100$ MeV in its inner regions. 
Thus, matter has $\mu_n/T \sim 1$, and so is neither degenerate nor
non-degenerate. Under such conditions numerical evaluation of the phase-space 
integrals in Eq.~(\ref{eq:full}) is necessary. 

This is done using an $NN$ amplitude reconstructed from the
experimental $NN$
phase shifts~\cite{SAID}. Although there is data on $nn$ scattering
only at very low energies, the wealth of experimental $pp$ data 
may be used to infer $T=1$ phase shifts. 
By isospin symmetry, these phase shifts 
should be the same as those for $nn$ scattering, 
up to small corrections for charge-independence breaking. 
Once a partial-wave decomposition is performed on ${\cal A}$ the $\phi$ 
integral can be performed analytically, leaving us to evaluate
a five-dimensional integral numerically. 
The results for the emissivity in neutron matter at
nuclear matter density, $n_N=n_0 \equiv 0.16~{\rm fm}^{-3}$,
in the case of two GODs
of radius 1 mm are represented by the solid line in Fig.~\ref{fig-emiss}.

%
\begin{figure}[h,t,b,p]
\centerline{{\epsfxsize=5.0in\epsfysize=3.5in\epsfbox{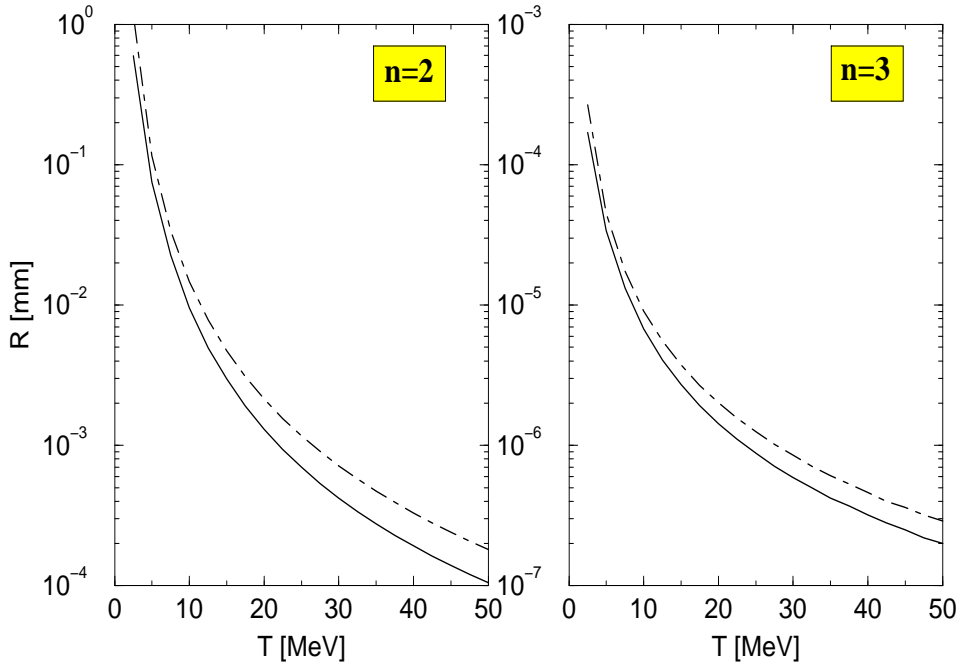}} }
\noindent
\caption{\it Bounds on the radius of the extra dimensions as a
function of temperature for neutron matter at $n_N=0.16~{\rm fm}^{-3}$
(solid line) and $n_N=0.48~{\rm fm}^{-3}$ (dot-dashed line).
The left panel gives the results for $n=2$ and the right panel
those for $n=3$.}
\label{fig-Rbounds}
\vskip .2in
\end{figure}

With these numerical results in hand we use the simplistic 
upper bound on the power per unit volume lost from SN1987a of
$\dot{\cal E}_{nn}~=~10^{19}~{\rm ergs/g/s}$~\cite{Raffelt} 
to place a bound on the radius $R$ of the extra dimensions. 
The resultant bounds are shown, as a
function of temperature for the cases $n=2$ and $n=3$, in
Fig.~\ref{fig-Rbounds}. The solid line is the result at
$n_N=0.16~{\rm fm}^{-3}$, while the dot-dashed line is the result
at three times nuclear matter density.

Assuming that our analysis applies to SN1987a,
and using $T~=~30~{\rm MeV}$ and a neutron number density 
equal to the density of nuclear matter,
our ``best" bound on the compactification radius is:
\begin{eqnarray}
R & < & 7.1 \times 10^{-4}~{\rm mm}
\ \ \ \  {\rm for}\ \ \ \ n=2,
\nonumber\\
R & < & 8.5 \times 10^{-7}~{\rm mm} \ \ \ \ {\rm for}\ \ \ \ n=3.
\label{eq:bound}
\end{eqnarray}
These bounds are a factor of two to three
weaker than those obtained by Cullen and Perelstein,
who found~\cite{CP}:
\begin{eqnarray}
R & < & 3 \times 10^{-4}~{\rm mm}
\ \ \ \  {\rm for}\ \ \ \ n=2,
\nonumber\\
R & < & 4 \times 10^{-7}~{\rm mm} \ \ \ \ {\rm for}\ \ \ \ n=3.
\label{eq:CPbound}
\end{eqnarray}
The source of the difference between the bounds (\ref{eq:CPbound})
and our bound (\ref{eq:bound}) will be discussed in Section~\ref{sec-discuss}.

\begin{figure}[t]
\centerline{{\epsfxsize=5.0in\epsfbox{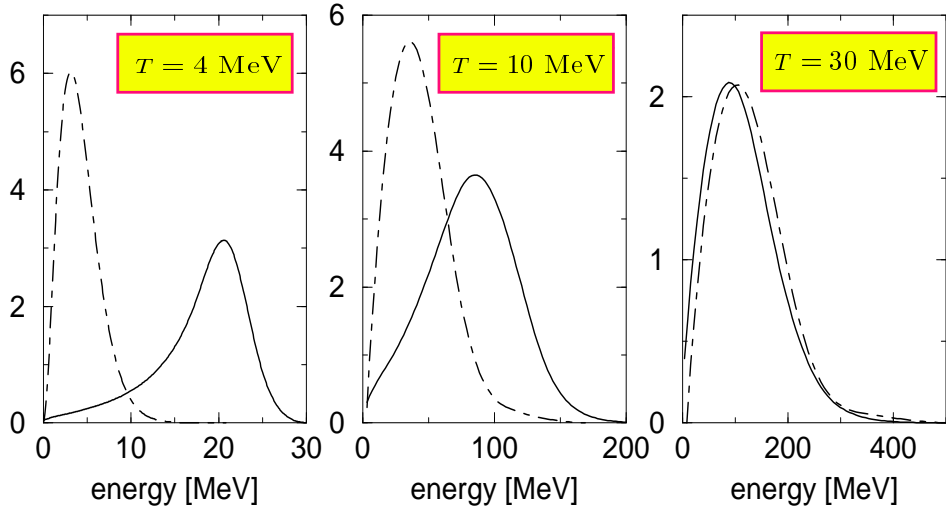}} }
\noindent
\caption{\it Two different partial
integrations of the dimensionless kernel, 
$K^{(n)}\left({\omega\over T}, {E^{\rm final}_{\rm cm}\over 2 T}\right)$,
for the case $n=2$, and nuclear matter density.
The solid curve corresponds to the final-state 
$nn$ energy distribution,
while the dot-dashed curve corresponds to the radiated particle
energy distribution.
For $T=10~{\rm MeV}$ and $T=30~{\rm MeV}$
the curves have been multiplied by a factor of
$10$ and $100$, respectively.
}
\label{fig-EXdist}
\vskip .2in
\end{figure}

It is interesting to examine the energy distribution of both the 
radiated particles and that of the emitting
nucleons, in order to determine if the soft-radiation approximation
is valid. To do this, note that the kernel $K^{(n)}(\alpha,u_r')$,
defined in Eq.~(\ref{eq:kernel}),
may be integrated over $u_r'$ or $\alpha$ in order to generate
a distribution in the radiation-energy variable or the nucleon-energy
variable. (Recall that $\alpha=\omega/T$ 
and $u_r'=E_{\rm cm}^{\rm final}/(2 T)$.) The results
of this procedure at three
different temperatures are displayed in Fig.~\ref{fig-EXdist},
as a function of the energy of the radiation and the final-state
energy of the nucleons.

%

Figure~\ref{fig-EXdist} shows that at low temperature the typical
radiated energy is much less than the typical total energy of the
final-state neutrons, and so soft-radiation theorems should work very
well. This is because in the degenerate limit the nucleon energy is
set by $p_F$, while the radiation's energy is proportional to
$T$. However, as the temperature is increased, and the neutrons become
non-degenerate, both energies scale with $T$. At temperatures relevant
to SN1987a the radiation is seen to carry off energy comparable to
that of the final-state nucleons. Note that if the neutron matter were
denser then the Fermi momentum of the nucleons would be higher, and so
the temperature where the peaks of the energy distributions coincide
would be somewhat larger.  On the other hand, as the number of extra
dimensions rises the power of $\alpha$ appearing in the kernel
$K^{(n)}$ increases, and so the radiation would tend to carry away even
more of the $NN$ state's energy.

The right-most panel of Fig.~\ref{fig-EXdist} represents the
conditions used in deriving the radius bounds (\ref{eq:bound}).  We
infer that roughly half of the initial energy of a typical $NN$ pair
is radiated away to the GODs. In other words, if gravitational radiation was
emitted into extra dimensions during SN1987a it may not have been
soft. The parameter $\chi$, defined in Eq.~(\ref{eq:chidefn}), which we
expect will control the expansion is of order 2/3 for the conditions
of interest. This is a rather large value for our ``small" parameter.
However, the first non-vanishing correction to the emissivity
is suppressed by two powers of $\chi$, which is somewhat reassuring.

\section{Limiting cases and analytic formulae}

\label{sec-analytic}

While the core region of SN1987a is neither degenerate nor
non-degenerate neutron matter~\cite{BT}, it is illuminating to
consider dilastrahlung and gravistrahlung processes in these extreme
cases.  As we shall show in this section, in both limits analytic
formulae for the emissivity in terms of one $NN$ cross section can be
obtained, provided certain observations about the differential
cross section are approximately true. Alternatively, effective-range
theory can be used to provide a form for $\frac{d\sigma}{d\Omega}$
which can be integrated analytically. In a supernova the neutron
momentum distribution lies essentially mid-way between degenerate and
non-degenerate, and so these limiting cases provide bounds on the true
emissivity.

\subsection{A non-degenerate neutron gas}

To analyze the non-degenerate limit of neutron matter the final-state
blocking factors $(1-f_i^\prime)$, are set equal to unity.  In other
words, all of the final-state phase space is open to the scattered
neutrons. In addition, the initial-state Fermi distributions are
replaced by Maxwell-Boltzmann distributions.  These simplifications
allow the emissivity to be written in terms of integrals over the
initial centre-of-mass energy distribution, the energy of the KK-dilaton
or KK-graviton, and the scattering angle
\begin{eqnarray}
{d{\cal E}_{nn}\over dt}
&=&  \frac{8 G_N}{3} \sqrt{\frac{T^7}{M}} n_N^2 \, \left( RT\right)^n h_n 
\, \pi^{n+1\over 2} 
\frac{\Gamma(n+1)}{\Gamma\left(n+\frac{5}{2}\right)}
\left[\frac{n^2 + 9n + 23}{(2n+7)(2n+5)}\right]
\nonumber\\
&& \qquad \int_0^\infty d\tau \, \tau^{n+4} e^{-\tau}
\int_{-1}^1 d(\cos \theta_{\rm cm}) \, \sin^2 \theta_{\rm cm} \,
{d\sigma\over d\Omega_{\rm cm}} (T \tau,\theta_{\rm cm}).
\label{eq:nondegem}
\end{eqnarray}
The variable $\tau=E_{\rm cm}^{\rm initial}/T$ is the ratio of the initial
$nn$ centre-of-mass kinetic energy 
to the temperature ($\tau=2u_r$), 
and $n_N$ is the number density of neutrons.
We have also used the relationship between ${\cal A}$ and the $NN$
differential cross section:

\begin{equation}
\frac{d \sigma}{d \Omega}=\frac{M^2 \pi^4 |{\cal A}|^2}{(2 \pi)^6}.
\end{equation}

In order to get the form (\ref{eq:nondegem}) we had to replace the
energy at which $\cal A$ is evaluated, $2 T \bar{u}$, by $2 T u$, the
initial-state energy. As we shall see below, this replacement makes
very little difference to the final result for $\dot{\cal E}_{nn}$. In
contrast, if we had employed the initial-state $NN$ energy $p^2/M$,
instead of the average energy $\overline{p^2}/M$, in the pre-factor
for the energy-loss rates $\frac{d \varepsilon_X}{d\omega}$, 
then the emissivity would have been larger
by roughly a factor of two.  However, as discussed in
Sections~\ref{sec-LO} and \ref{sec-NLO}, the choice $\overline{p^2}/M$
is the one that pushes the error in the LO result to $O(\chi^2)$.
Therefore, the formula (\ref{eq:nondegem}) is an excellent
approximation to the NLO result for the emissivity in the
non-degenerate limit.

We now consider two analytic forms for the $nn$ scattering cross section.

\subsubsection{Nondegenerate neutron gas: 
energy-independent ${d\sigma\over d\Omega}$}

The $nn$ differential cross section is, in fact, essentially
energy-independent in the region of interest. The angular dependence
is also weak, except at small angles, and the contribution 
of these angular regions is
suppressed by the presence of $\sin^2 \theta_{\rm cm}$.
The emissivity can thus be estimated as 
\begin{eqnarray}
{d{\cal E}_{nn}\over dt}
& = & 
\frac{16 G_N}{9} \sqrt{\frac{T^7}{M}}
\left( RT\right)^n\ \pi^{{n-1\over 2}}  h_n\ n_N^2\ \sigma^{nn}_0\ 
{\Gamma\left(n+1\right) \Gamma(n+5)
\over\Gamma\left(n+\frac{5}{2}\right)} \frac{n^2 + 9n + 23}{(2n+7)(2n+5)}.
\label{eq:ndrough}
\end{eqnarray}
The $nn$ total cross section can be extracted from $pp$ scattering data,
provided that Coulomb effects are removed. Doing this, and evaluating
$\sigma^{nn}_0$ at the peak of 
the function $\tau^{n+4} e^{-\tau}$,
leads to $\sigma^{nn}_0=25~{\rm mb}$, which in turn yields bounds:
\begin{eqnarray}
R & < & 7.0 \times 10^{-4}~{\rm mm}
\ \ \ \  {\rm for}\ \ \ \ n=2,
\nonumber\\
R & < & 8.3 \times 10^{-7}~{\rm mm} \ \ \ \ {\rm for}\ \ \ \ n=3,
\label{eq:boundeind}
\end{eqnarray}
at $T=30$ MeV and $n_N=n_0$.
These limits are very close to the ``best" bounds
(\ref{eq:bound}). This indicates that, under the conditions
examined here, the neutron gas is very nearly non-degenerate,
and the neutrons scatter via a cross section which is 
almost independent of angle and energy. Note that even had
the $nn$ cross section not been so flat the result (\ref{eq:boundeind})
could still be taken as a result of naive dimensional analysis,
since any ``typical" $NN$ cross section will give numbers of this
order of magnitude. Its success as such an estimate hinges 
on the fact that
Eq.~(\ref{eq:ndrough}) approximates the temperature dependence of the 
full result very well (see Fig.~\ref{fig-emiss}).

\subsubsection{Nondegenerate neutron gas: 
${d\sigma\over d\Omega}$ From Effective-range theory}

Effective-range theory (ER) describes very low-energy 
nucleon-nucleon scattering by a unique momentum expansion
about $|\vec{p}^{\,}|=0$~\cite{ERtheory}. Empirically,
the first two terms in the 
expansion of $|\vec{p}^{\,}| \cot \delta$ describe the 
S-wave cross section well, 
even at energies greater than the naive breakdown
scale of the expansion, which is $|\vec{p}^{\,}|=m_\pi/2$.
The $nn$ non-relativistic differential cross section in ER is,
neglecting contributions beyond the effective range,
\begin{eqnarray}
{{d\sigma^{nn}_{ER}}\over {d\Omega_{\rm cm}}} & = & 
\left|{1\over -{1\over a_0} + {1\over 2} r_0 |\vec{p}^{\,}|^2\ 
- \ i  |\vec{p}^{\,}|} \right|^2
\nonumber\\
& \rightarrow &
{1\over |\vec{p}^{\,}|^2
\left( 1 + {1\over 4} r_0^2 \, |\vec{p}^{\,}|^2\right)}
\ \ \ {\rm for}\ \ \  |\vec{p}^{\,}| \gg {1\over a_0},
\label{eq:ER}
\end{eqnarray}
where $a_0=-18.5\pm 0.4~{\rm fm}$ 
and $r_0=2.80\pm 0.11~{\rm fm}$ 
are the scattering length and effective range for $nn$ scattering
in the $^1S_0$ channel~\cite{nndata}.
As the typical momentum involved in the scattering events of interest is
much larger than the inverse scattering length 
we take the second form for the differential cross section.
Inserting this in Eq.~(\ref{eq:nondegem})
leads to a simple expression for the emissivity:

\begin{eqnarray}
{d{\cal E}^{\rm ER}_{nn}\over dt} =  
\frac{1024 G_N}{9} \left( RT\right)^n\  && h_n \pi^{1+n\over 2}\ 
n_N^2 {\Gamma\left(n+1\right) \Gamma(n+4) 
\over M^4 r_0^5  \Gamma\left(n+\frac{5}{2}\right)}\nonumber\\
&& \frac{n^2 + 9n +23}{(2n+7)(2n+5)}
\ x^{n+{3\over 2}}
e^{x} \,  \Gamma\left( -(n+3), x\right),
\label{eq:dger}
\end{eqnarray}
with:
\begin{equation}
x  =  {4\over M T r_0^2},
\end{equation}
and $\Gamma\left(x,y\right)$ the incomplete Gamma function,
which obeys $\Gamma\left(z,0\right)=\Gamma\left(z\right)$.
The bounds obtained from the ``Raffelt criterion"
$\dot{\cal E} \leq 10^{19}$ ergs/g/s at $n_N=n_0$
and $T=30$ MeV are
\begin{eqnarray}
R & < & 2.4\times 10^{-3}~{\rm mm}
\ \ \ \  {\rm for}\ \ \ \ n=2,
\nonumber\\
R & < & 2.1 \times 10^{-6}~{\rm mm} \ \ \ \ {\rm for}\ \ \ \ n=3.
\label{eq:bounder}
\end{eqnarray}

These bounds are much weaker than the bounds (\ref{eq:boundeind}) and 
(\ref{eq:bound}). As expected, they give an upper bound on $R$, since
the cross section (\ref{eq:ER}) includes only S-waves, and other
partial waves also contribute significantly to $\sigma_0^{nn}$ 
in the energy regime of interest.

\subsection{A degenerate neutron gas}

Supernovae are, of course, systems far from the degenerate limit.
However, the degenerate limit is of particular interest for
neutron-star cooling, as stressed by Friman and Maxwell~\cite{FM} and
Iwamoto~\cite{Iwamoto} in their respective studies of $\nu \bar{\nu}$
and axion radiation. In this section we employ some of the techniques
used in Refs.~\cite{FM,Iwamoto} to derive analytic formulae for this case,
where $\mu/T \gg 1$. 

Radiation from a degenerate neutron gas 
can only arise from scatterings involving
neutrons near the Fermi surface in both the 
initial and final states.
In the soft-radiation limit, the momentum of each nucleon involved in the 
scattering event is constrained to lie on the Fermi surface.
This approximation leaves only angular and energy integrals 
in the calculation of the emissivity.
The energy integrals are those that enter Fermi liquid theory,
and are well known~\cite{BP}. In particular,
the energy integral that occurs in the case of $2\rightarrow 2$ scattering is
\begin{eqnarray}
\int dE_1 \, dE_2 \, dE_1' \, dE_2'\  f_1 f_2 (1-f_1') (1-f_2')
\delta(E_1 + E_2 + E_1' + E_2' - \omega)
& = &
{1\over 6} {\omega (\omega^2+4\pi^2 T^2)\over e^{\omega/T} - 1}.
\label{eq:ff}
\end{eqnarray}
The $\omega$ integral can then be done analytically, leaving
two angular integrations, and a result, for a degenerate neutron
gas, of:
\begin{eqnarray}
{d{\cal E}_{nn}\over dt}
& = & 
G_N \left( RT\right)^n\  h_n\ Y_n\ 
{ T^4\  \Gamma\left(n+4\right)\  \pi^{{n-15\over 2}}\over 
36 }
\int\  dt \, d\phi \,
\ |\vec{p}^{\,}_1-\vec{p}^{\,}_2|^4\  {d\sigma\over d\Omega_{\rm cm}} 
\ \sin^2\theta_{\rm cm}
\nonumber\\
Y_n & = & \zeta (n+4)\
\ +\ 
4\pi^2 {\zeta (n+2)\over (n+2)(n+3)},
\label{eq:emdeg}
\end{eqnarray}
where $\zeta (x)$ is the Riemann Zeta function.
The kinematic variables in the centre-of-mass frame are related to the 
integration variables $t$ and $\phi$,
defined in the rest frame of the Fermi gas by
\begin{eqnarray}
\cos\theta_{\rm cm} & = & 1-{2 t^2\over |\vec{p}^{\,}_1-\vec{p}^{\,}_2|^2},
\nonumber\\
|\vec{p}^{\,}_1-\vec{p}^{\,}_2|^2 & = & 2 p_F^2 \left[ 1 + {t^2\over 4 p_F^2}
\ -\ \left( 1 - {t^2\over 4 p_F^2}\right)\cos\phi\right],
\label{eq:kinrel}
\end{eqnarray}
where $p_F$ is the Fermi momentum, related to the neutron number density
by $n_N=p_F^3/(3\pi^2)$. Note that this method of calculation is only 
valid to LO in the soft expansion, since we have implicitly assumed
that all nucleon momenta are on the Fermi surface.

Once again, we can now consider two approximate forms for the $nn$
scattering cross section. Firstly, evaluation of the integral appearing in 
Eq.~(\ref{eq:emdeg}) is straightforward if the 
differential cross section is independent of energy and angle.
We find that
\begin{eqnarray}
{d{\cal E}_{nn}\over dt}
& = & 
G_N\left( RT\right)^n\  h_n\ Y_n\ \sigma^{nn}_0\ 
{16\  T^4\   \pi^{{n-15\over 2}}\  \Gamma\left(n+4\right) \ p_F^5
\over 135}.
\label{eq:degrough}
\end{eqnarray}

On the other hand, explicit evaluation of Eq.~(\ref{eq:emdeg}) with the 
differential cross section of ER, Eq.~(\ref{eq:ER}), yields
\begin{eqnarray}
{d{\cal E}^{\rm ER}_{nn}\over dt}
& = & 
G_N\left( RT\right)^n\  h_n\ Y_n\ 
{4 \  T^4\  \pi^{n-13 \over 2} \ \Gamma\left(n+4\right) \ p_F^3 
\over 9 }
\ \ D^{({\rm ER})}\left({p_F r_0\over 2}\right),
\nonumber\\
D^{({\rm ER})}\left(x\right) & = & 
{1\over x^2}\ -\ {\log\left[ x + \sqrt{1+x^2}\right]\over x^3 \sqrt{1+x^2}}.
\label{eq:ndeger}
\end{eqnarray}

We will not use these formulae to place bounds on $R$ from SN1987a.
The nucleonic gas of SN1987a was far from degenerate, and so 
such bounds are not particularly meaningful.

\section{Discussion}

\label{sec-discuss}

\begin{figure}[t]
\centerline{{\epsfysize=3.5in\epsfbox{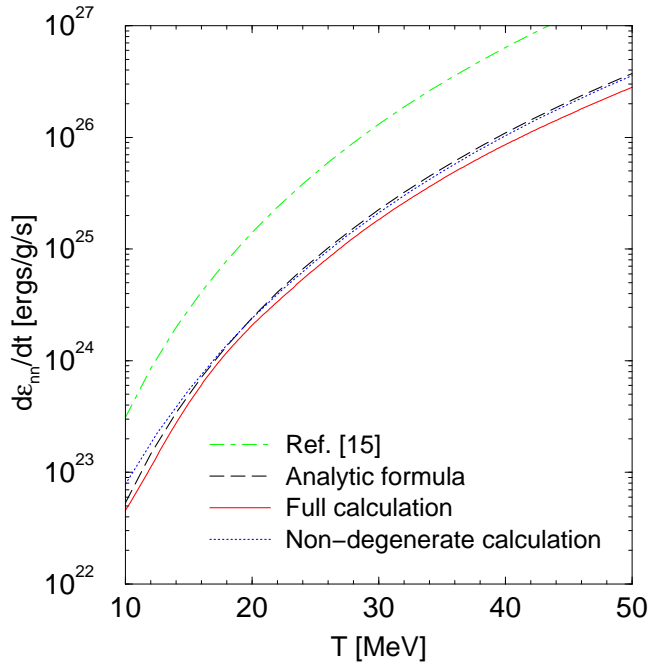}} }
\noindent
\caption{\it Emissivity of SN1987a due to the reactions $\dila$ and $\grav$,
calculated assuming there are two GODs of radius $R=1$ mm. The 
solid line is the full calculation, employing Eq.~(\ref{eq:full}),
while the dotted and dashed curves represent the
different approximations to this discussed in the text. 
The dot-dashed line is the result of Cullen and Perelstein, 
Ref.~\protect\cite{CP}.}
\label{fig-emiss}
\vskip .2in
\end{figure}

In Figure~\ref{fig-emiss} we present numerical results for the
KK-graviton emissivity of a neutron gas, due to neutron-neutron
collisions in the soft-radiation limit, at nuclear matter density.
The solid curve shows numerical results obtained by using the
``measured", angle-dependent, elastic neutron-neutron differential
cross section and the full Fermi distributions for the initial- and
final-state nucleons.  This numerical result, which is exact at the
two-body level, provided the radiation is soft, is to be compared with
the result given by Eq.~(\ref{eq:full}) when the final-state
blocking is ignored and the initial-state Fermi-Dirac distributions
are replaced by Maxwell-Boltzmann distributions. This result, which
is valid in the non-degenerate limit, is represented by the dotted curve.  
The dashed line is the analytic result
(\ref{eq:ndrough}), that is obtained by neglecting the energy and
angular dependence of
the $nn$ differential cross section. This function depends only on the
``measured" total $nn$ cross section. Finally, the dot-dashed line is the
result quoted by Cullen and Perelstein for $n=2$ in Ref.~\cite{CP}:

\begin{equation} 
\frac{d {\cal E}_{nn}}{dt}=
3.7 \times 10^{17}~\rm{ergs/g/s} \ \rho_{14} T_{\rm MeV}^{5.5} 
R_{\rm mm}^2,
\label{eq:CPemiss}
\end{equation} 
where $\rho_{14}$ is the mass density in units of $10^{14}~{\rm g/cm^3}$.
Eq.~(\ref{eq:CPemiss}) was derived using the one-pion-exchange
approximation for the matrix element and the non-degenerate limit for
the neutron gas.

%

Figure~\ref{fig-emiss} shows that the conditions examined here are
close to the non-degenerate limit. Calculations employing this limit produce
emissivities only 15\% larger than the full calculation using
Eq.~(\ref{eq:full}). Furthermore, the approximation that the $NN$
cross section is flat in both energy and angle works exceptionally
well. 
The effective-range theory result---which is not displayed in 
Fig.~\ref{fig-emiss}---has the wrong temperature dependence. The
``full" temperature dependence---at least above $T \approx 10$ MeV---is
very close to the result from a flat cross section and
the non-degenerate limit: $\dot{\cal E} \, \alpha \, T^{5.5}$.
The emissivity increases with temperature more slowly than this if 
the effective-range theory is employed because the 
cross section (\ref{eq:ER}) drops with increasing $nn$ energy.

%
\begin{figure}[h,t,b]
\centerline{{\epsfysize=5.0in\epsfbox{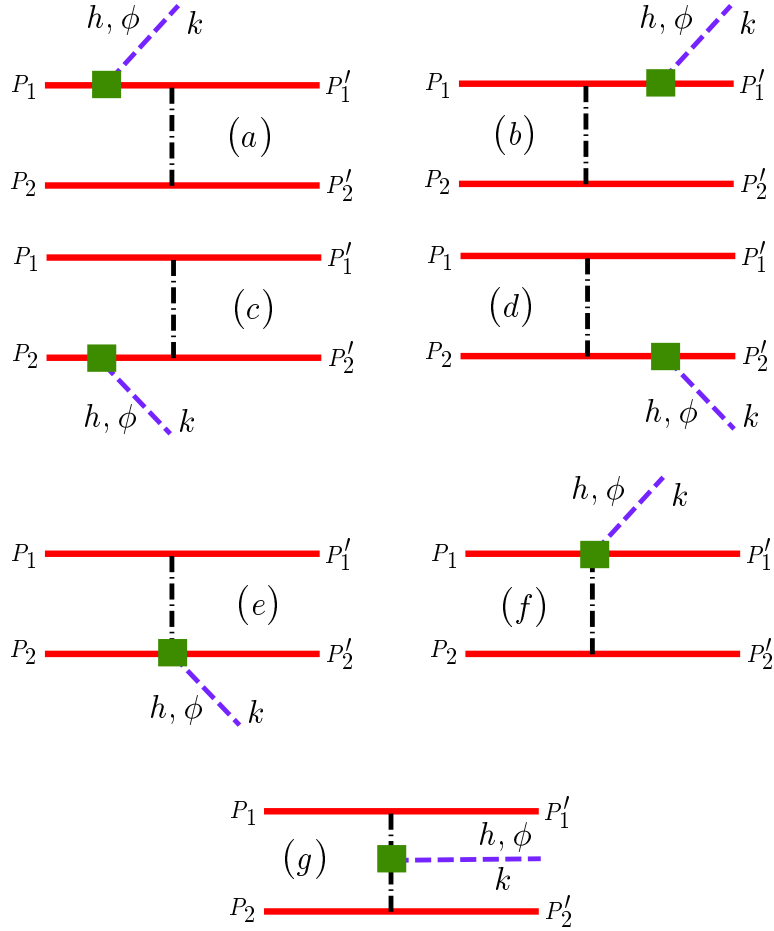}} }
\noindent
\caption{\it The seven graphs, which, together with their fermion interchange
partners, were calculated to obtain the KK-graviton and KK-dilaton
emissivity in previous works. The solid lines are nucleons,
the dashed line is the emitted KK-graviton or KK-dilaton, and
the dot-dashed line represents the pion.}
\label{fig-OPEgraphs}
\vskip .2in
\end{figure}

The temperature dependence $\dot{\cal E} \, \alpha \, T^{5.5}$ is the
same as that found in Refs.~\cite{CP,BHKZ}, although
Fig.~\ref{fig-emiss} shows that the overall size of the emissivity
(\ref{eq:CPemiss}) is significantly larger.  Refs.~\cite{CP,BHKZ} both
calculated KK-graviton radiation from the $NN$ system by coupling
gravitons in all possible ways to the one-pion-exchange graph. The
resultant diagrams are depicted in Fig.~\ref{fig-OPEgraphs}.  The
authors of both Refs.~\cite{CP,BHKZ} had graphs (a)-(d) giving zero,
with graphs (e)-(g) supplying the full result. This apparently occurs
because the square of the matrix element was only calculated up to
order $p^2/M^2$, and graphs (a)-(d) do not contribute until $O(p^4)$.
If diagrams (e)-(g) are the full amplitude for $\grav$ and $\dila$
then, in the limit $m_\pi^2 \ll M T$ the matrix element is constant
with $\omega$ and the $NN$ system's energy\footnote{Replacing the full
one-pion-exchange-approximation matrix element by its value in the
$m_\pi \rightarrow 0$ limit can be a poor approximation. In the case
of axial radiation it enhances the emissivity by as much as a factor
of two, as discussed in Refs.~\cite{BT,RafOPE}.}.

Our work here shows that in the soft regime, where $\omega \ll p^2/M$,
graphs (a)-(d) will dominate over graphs (e)-(g). Consequently, if
Refs.~\cite{CP,BHKZ} had examined a degenerate system, where the
gravitational radiation is soft, they would have missed the dominant
contribution to the emissivity, and so obtained the wrong temperature
dependence. On the other hand, in the supernova case the matter is far from
being degenerate, and as the discussion below Fig.~\ref{fig-EXdist}
shows, the radiation is not especially soft. In fact, naive dimensional
analysis shows that in the non-degenerate limit:

\begin{equation}
\frac{p^4}{M^2 \omega^2} \sim T^0.
\end{equation}
So, it should come as no surprise that our graphs, which represent
radiation from external legs, give the same temperature dependence
as graphs (e)-(g) of Fig.~\ref{fig-OPEgraphs}.
Graphs (e)-(g)
correspond to radiation from {\it internal} lines, and 
appear first at $O(\chi^0)$ in our approach, but they will be 
accompanied there by many other diagrams. One group of additional graphs 
at $O(\chi^0)$ will
represent the gravitational
radiation coupling to short-distance components of the $NN$ force,
while another group involves the inclusion of final and initial-state
interactions for the neutrons. Both of these effects should give sizeable
corrections to graphs (e)-(g), and neither was
included in the calculations that led to the result (\ref{eq:CPemiss}).

Thus, we would argue that the result (\ref{eq:CPemiss})
is best viewed as naive
dimensional analysis of the emissivity $\dot{\cal E}_{nn}$ due to
$\grav$. In this context, we note that
either Eq.~(\ref{eq:ndrough}) or the graphs
(e)-(g) of Fig.~\ref{fig-OPEgraphs} can be used to estimate the
emissivity for a non-degenerate gas of nucleons, and the dependence 
they predict on
the temperature, and on the size and number of GODs, will be correct. 
The coefficient multiplying the overall result will 
differ, but it will involve ``typical nuclear scales" and
so it is, perhaps, not surprising that the resultant emissivities are
within an order of magnitude of each other. Ultimately though,
our result for $\dot{\cal E}_{nn}$ is significantly smaller than
that of Refs.~\cite{CP,BHKZ}, and our bounds on $R$ are 
correspondingly weaker.

The bounds (\ref{eq:bound}) obtained in this work on the type
of extra dimensions proposed in Refs.~\cite{ADD1,ADD2,AADD} are
model-independent, and are valid under the assumptions that the
radiation is soft and that two-body collisions dominate the
emission. It appears that conditions in SN1987a are such that
radiation from the $NN$ system is not soft enough to make the
calculation we have presented here completely
reliable. However, the framework described in this paper holds out the
promise that corrections to our result can be
systematically computed.  In this way, it represents a significant
advance on previous work, where uncontrolled approximations for the
$NN$ dynamics were employed.  Some of the higher-order corrections to the NLO
result which was used to set the bounds (\ref{eq:bound}) can still be
expressed in terms of $NN$ system observables,  e.g. derivatives of the $NN$
differential cross section with respect to angle and energy. Indeed,
since the $NN$ differential cross section is essentially constant with
energy and angle in the region probed by our calculation
one might hope that the $O(\chi^0)$ contribution to $\dot{\cal E}$
is fairly small.  Even were higher-order corrections to halve the
emissivity the bound on the radius for the $n=2$ case
would still be a severe $R \ltap 1 \times 10^{-3}$ mm. This is two
orders of magnitude beyond the bound attainable at any existing
collider~\cite{colliderKK}. The only comparably stringent bound on the
radius of toroidally-compactified, large, GODs comes from the
cosmological considerations discussed by Hall and Smith in
Ref.~\cite{HS}, who advocate $R < 5.1 \times 10^{-5}$ mm. However, as
Hall and Smith themselves comment, this bound is subject to their
ignorance of the effect that GODs have on cosmology.

In summary, for the case of two or three large, toroidally-compactified
``gravity-only" dimensions 
one of the best bounds on modifications in
the physics of gravity comes from delving into the
extreme conditions present in SN1987a, and asking what effects such
changes would have there. Our detailed examination of the two-nucleon
dynamics which yields this bound implies that the radius of these
extra dimensions must be below $7.1 \times 10^{-4}$ mm for 
$n=2$ and below $8.5 \times 10^{-7}$ mm for $n=3$.

\section*{Acknowledgments}
We thank George Bertsch and Zacharia Chacko for  
useful conversations, and Brad Keister
for drawing our attention to Ref.~\cite{AD}.
This work is supported in part by the U.S. Dept. of Energy under
Grants No. DE-FG03-97ER4014 and DOE-ER-40561. C.~H. acknowledges
the support of the Humboldt foundation.

\vfill
\eject

\end{document}